\documentclass{iopart}

\begin{document}

 \jl{6}
 
\title{Why do we observe a small but non zero  cosmological constant ?} 
\author{T. Padmanabhan \footnote[1]{E-mail address:  {\tt nabhan@iucaa.ernet.in}}}
\address{Inter-University Centre for Astronomy and Astrophysics, Post Bag 4, Ganeshkhind, Pune - 411 007, India.}

\begin{abstract}
   The current observations seem to suggest that the universe has a positive cosmological constant 
  of the order of $H_0^2$ while the most natural value for the cosmological constant will be $L_P^{-2}$ where
   $L_P = (G\hbar/c^3)^{1/2}$ is the Planck length. This reduction of the cosmological constant from $L_P^{-2} $ 
   to $L_P^{-2}(L_PH_0)^2$  may be interpreted as due to the ability of quantum micro structure of spacetime
   to readjust itself and absorb bulk vacuum energy densities. Being a quantum mechanical process, such a 
   cancellation cannot be exact and the residual quantum fluctuations appear as the ``small'' cosmological constant.
   I describe the features of a toy model for the spacetime micro structure which could allow for the bulk vacuum 
   energy densities to be canceled leaving behind a small residual value of the the correct magnitude. 
   Some other models (like the ones  based on canonical ensemble for the four volume or quantum fluctuations of the
   horizon size) lead to an insignificantly small value of $H_0^2(L_PH_0)^n$ with $n=0.5-1$ showing that obtaining 
   the correct order of magnitude for the residual fluctuations in the cosmological constant is a nontrivial task, 
   becaue of the existence of the small dimensionless number $H_0L_P$ . 
\end{abstract}

\maketitle 

\noindent
The action for classical gravitational field depends on the speed of light,
$c$, the Newtonian gravitational constant, $G$ and the cosmological
constant, $\Lambda$. Since it is not possible to produce a dimensionless number
from these three constants, their relative values have no meaning and with a 
suitable choice of units we can set all the three of them to unity if they are nonzero and positive,
  say. The situation 
is different in quantum theory which introduces the constant $\hbar$.
  It is then possible to form the dimensionless number $\Lambda (G\hbar/c^3) \equiv \Lambda 
  L_P^2 $ where $L_P\approx 10^{-33} $  cm is the Planck length. 
  If we assume that dimensionless combinations of coupling constants should be of order unity, then the natural value for cosmological constant will be $\Lambda\approx L_P^{-2}$.
  Current cosmological observations, (e.g., \cite{sperl}  and \cite{agr}), however, suggest that, 
 the {\it effective value} value of $\Lambda $ 
    (which will pick up contributions from all vacuum energy densities of matter fields) has been 
    reduced from the natural value of  $L_P^{-2}$ to $L_P^{-2}(L_PH_0)^2$ where $H_0$
    is the current value of the Hubble constant.
    If these observations are correct, then we need to answer two separate questions: (i) Why does
     large amount
    of vacuum energy density remain unobservable by gravitational effects ?  (ii) Why does
     a very tiny part of it
     appears as observable  residue ? [Attempt to understand the nature of the cosmological constant has a long history. Some of the pioneering ideas in this subject are by Zeldovich \cite{yz}; also see \cite{dolgov}.
For a review of issues related to 
     cosmological constant, see \cite{carol}  and references cited therein.]
    
    An attractive way of thinking about these questions is the following: Let us assume that the quantum
    micro structure of spacetime at Planck scale is capable of readjusting itself, soaking up any
    vacuum energy density which is introduced ---  like a sponge soaking up water.
    If this process is fully deterministic and exact, all vacuum energy densities will cease to have
    macroscopic gravitational effects. However, since this process is inherently quantum gravitational,
    it is subject to quantum fluctuations at Planck scales. 
    Hence, a tiny part of the vacuum energy will survive the 
    process and will lead to observable effects.      
    I would conjecture that the cosmological constant we measure corresponds to this  small 
    residual fluctuation which will depend on the volume of the spacetime region that is probed.  
    It is small, in the sense that it has been reduced from $L_P^{-2}$ to $L_P^{-2}(L_PH_0)^2$, 
    which indicates the fact that fluctuations --- when measured over a large volume --- is small
     compared to the bulk value. It is the wetness of the sponge we notice, not the water content inside. 
     
     To make further progress with such an idea, one needs to know the exact description of spacetime
     micro structure in a quantum theory of gravity. Since this is not available, I will proceed in a more 
     tentative and speculative manner illustrating the idea in three stages. To begin with, I will provide
     a description of spacetime micro structure in which cancellation of bulk of the vacuum energy 
     density is indeed possible. Next, I will explore the consequences of such a model in the 
     semi classical limit. Finally, I will indicate how the residual fluctuations can survive after the 
     cancellation of the bulk vacuum energy and provide the cosmological constant of the same order
     as observed in the universe. 
     The first part is an excursion into unknown territory and is necessarily speculative.  The second part is simple and
     rigorous while the third part is fairly straightforward in the context of quantum cosmology.
     Part of the motivation in presenting these ideas is to generate further interest in this approach
     so that better models can be constructed.
     
     Let me begin by asking how the action in 
     Einstein's gravity arises in the long wavelength limit of some (unspecified) description of spacetime
     micro structure. If the macroscopic spacetime is divided into proper four volumes of size
     $(\Delta x)^4$ then the long wavelength limit will correspond to $(\Delta x/L_P)^4 \gg 1$. 
     In the classical limit, we will let $(\Delta x)^4$ to be replaced by the integration element
     for the proper 4-volume $\sqrt{-g}\, d^4 x$. [The description is similar to the one 
     used in kinetic theory of gases in which a spatial volume $d^3{\bf x}$ will be treated 
     as infinitesimal for the purposes of calculus but --- at the same time --- is expected to contain 
     sufficiently large number of molecules in order to provide a smooth fluid approximation.]
      Given a large 4-volume ${\cal V}$ of the  spacetime, we will divide it into 
     $M$  cubes of size $(\Delta x)^4$ and label the cubes by $n = 1, 2,  ....., M$.
     The contribution to the path integral amplitude ${\cal A}$, describing long wavelength
     limit of gravity, can be expressed in the form 
     \begin{equation}
     {\cal A} = \prod_n \left[ 1+\left(c_1 (RL_P^2)+c_2 (RL_P^2)^2 + \cdots \right) {i(\Delta x)^4\over
     L_P^4}\right]
     \end{equation}
     The nature of the terms  within the brackets ( ) is essentially dictated by symmetry and dimensional
     considerations with $c_1, c_2$ etc. being numerical constants. The leading term
     should obviously be proportional to $R$ to reproduce Einstein's theory; but it is  possible to have non polynomial
     sub leading term like $\ln (R L_P^2)$, or even combinations involving other curvature 
     components. I will not be concerned with these terms (and have not explicitly shown them) since the classical 
     gravity only cares for the leading term. (This is obvious from the 
     facts that $L_P^2 \propto \hbar$ while the classical action should be independent
     of $\hbar$.) Also note that we have ignored a constant
     term --- which will represent the cosmological constant --- for the moment; this, of course,
     will be discussed in detail later on. Writing $(1+x) \approx e^x$, the amplitude becomes
     \begin{eqnarray}
     {\cal A} &=& \prod _n \left[ \exp (c_1 (RL_P^2)+\cdots )\right]^{{i(\Delta x)^4\over L_P^4}} \nonumber \\
    & \to &
     \exp {ic_1\over L_P^4} \int d^4x \sqrt{-g} (RL_P^2)\label{stdres}
     \end{eqnarray}
     where in the last equation I have indicated the standard continuum limit. (In conventional units
      $c_1 = (16\pi)^{-1}$.) So far, I have
     merely reinterpreted the conventional results.
     
     Let us now ask how one could describe the ability of spacetime micro structure to readjust
     itself and absorb vacuum energy densities. This would require some additional dynamical 
     degree of freedom that will appear in the path integral amplitude and survive in the 
     classical limit. Let us describe this feature by modifying the amplitude $[ \exp (c_1 (RL_P^2)+\cdots )]$
     in the above equation by a factor $[\phi(x_n)/\phi_0]$ where $\phi(x) $ is a scalar
     degree of freedom and $\phi_0$ is a pure number introduced to keep this factor
     dimensionless. In other words, I modify the amplitude to the form:
     \begin{equation}
     {\cal A}_{\rm modify} = \prod_n \left[ {\phi(x_n)\over \phi_0} e^{[c_1RL_P^2+\cdots]}
     \right]^{{i(\Delta x)^4\over L_P^4}}\label{modres}
     \end{equation}
     Since this is the basic assumption which I have introduced, let us pause for a moment
     to discuss it. 
     
     Different approaches to quantum gravity have different descriptions of
     microscopic spacetime structure (strings, loops, ...); however, all of them need to reproduce
     the action $A_{\rm gr}$ for classical Einstein gravity in the long wavelength limit, which
     is equivalent to providing a path integral amplitude $\exp(iA_{\rm gr})$ in the semiclassical
     limit. Since $A_{\rm gr}$ is expressible as a spacetime integral over a local Lagrangian
     density, all these approaches will lead to something similar to equation (\ref{stdres})
     at the appropriate limit. If this is the whole story, no trace of quantum gravitational micro structure
     survives at macroscopic scales and I cannot implement the basic paradigm of macroscopic
     vacuum energy densities being compensated by readjustment of spacetime micro structure.
     Equation (\ref{modres}), on the other hand, states that the correct approach to 
     quantum gravity will lead to the survival of an extra bulk degree of freedom [denoted by $\phi(x)$]
     which characterizes the ability of each Planck scale volume element to readjust to 
     vacuum energy densities. This modification is certainly the simplest possible one mathematically.
     
     Let us see how this works in the long wavelength limit.
     The extra factor in (\ref{modres}) will lead to the term of the form 
     \begin{eqnarray}
     \prod_n \left({\phi\over \phi_0}\right)^{{i(\Delta x)^4\over L_P^4}}
     &=&\prod_n \exp \left[{{i(\Delta x)^4\over L_P^4}}
     \ln \left({\phi\over \phi_0}\right)\right]\nonumber \\
     & \to & 
     \exp {i\over L_P^4} \int d^4 x \sqrt{-g} \ln \left({\phi\over \phi_0}\right)
     \label{phifield}
     \end{eqnarray}
     Thus, the net effect of our assumption is to introduce a `scalar field  potential' $V(\phi) = 
     - L_P^{-4} \ln \left(\phi/ \phi_0\right)$ in the semiclassical limit. It is obvious that
     the rescaling of such a scalar field by $\phi \to q \phi$ is equivalent to adding a cosmological
     constant with vacuum energy $-L_P^{-4} \ln q$. Alternatively, any vacuum energy
     can be re absorbed by such a rescaling. 
     
     I am {\it not} suggesting that $\phi$ is a fundamental scalar field with a logarithmic  potential; rather,
     it is a residual degree of freedom arising from unknown quantum micro structure of 
     spacetime and surviving to macroscopic scales.
     The difference between these two points of view is vital.
     To see this explicitly, consider a classical gravitational field coupled to 
      a scalar field with the action 
     \begin{eqnarray}
     A & = & {1\over 16 \pi L_P^2} \int (R- 2\Lambda) \sqrt{-g} d^4 x \nonumber \\
     & \quad +& \int \sqrt{-g} d^4 x \left[ {1\over 2} \phi^i
     \phi_i 
       + L_P^{-4} \ln \left({\phi\over \phi_0}\right)\right]\label{gravaction}
     \end{eqnarray}
     It may seem that 
     we can  absorb $\Lambda$  by a rescaling even now. Indeed, the
     action in (\ref{gravaction}) is invariant under the transformations 
     \begin{equation}
     \phi \to q\phi;\qquad  x^a \to f x^a;\qquad
     L^2_{P\, {\rm new}} = {L_P^2\over f^2} 
     \label{eqsix}
     \end{equation}
      with 
      \begin{equation}   
     q=\exp\left({\Lambda L_P^2\over 8\pi}\right); \qquad f = {1\over q}
     \end{equation}
      If the original cosmological constant was such that
     $\Lambda L_P^2 = {\cal O}(1)$, then $q$ and $f$ are order unity parameters and the renormalized
     value of the Newtonian constant differs from the original value only by a factor of order unity.
     If $\Lambda L_P^2 \ll 1$ the same result holds with greater accuracy. 
     The difficulty is that, if we treat $\phi$ as a {\it dynamical} field, then the term
     \begin{equation}
     A_0 = \int d^4 x \sqrt{-g} \left[ L_P^{-4} \ln {\phi\over \phi_0} - {\Lambda\over 8\pi L_P^2}\right]
     \label{anot}
     \end{equation}
     in the action will evolve and contribute a vacuum energy density of ${\cal O}(L_P^{-4})$ which, of course,
     we do not want. The fact that the scalar degree of freedom occurs as a potential in (\ref{phifield})
     without a corresponding kinetic energy term shows that its dynamics is unconventional and nonclassical.

     The above description in terms of macroscopic scalar degree of freedom can, of course, be only approximate.
      Treated as a  vestige
     of a quantum gravitational degrees of freedom, the cancellation in (\ref{anot}), leading to $A_0 =0$,
     cannot be
     precise because of fluctuations in the elementary spacetime volumes.
     These fluctuations will reappear as a ``small''  cosmological constant
     because of  two key ingredients:
 (i) discrete spacetime structure at Planck length and (ii) quantum gravitational
uncertainty principle. 

To show this, we first note that the net
cosmological constant can be thought of as a lagrange multiplier for proper volume
of spacetime in the action functional for gravity arising from the $A_0$ term in (\ref{anot}).
In any quantum cosmological models which leads to large volumes for the universe, phase of
the wave function will pick up a factor of the form
\begin{equation}
\Psi\propto \exp(-iA_0) \propto 
 \exp\left[ -i\left({\Lambda_{\rm eff}{\cal V}\over 8 \pi  L_P^2}\right)\right]
 \end{equation}
 from  (\ref{anot}), where ${\cal V}$ is the four volume. 
Treating $(\Lambda_{\rm eff}/8 \pi  L_P^2,{\cal V})$ as conjugate variables $(q,p)$, we can invoke the standard uncertainty principle to predict
 $\Delta\Lambda\approx
8 \pi  L_P^2/\Delta{\cal V}$. Now we use the earlier  assumption regarding the microscopic structure of the spacetime: Assume that there is a zero point length  of the order of $L_P$
so that the volume of the universe is made of a large number ($N$) of cells, each of volume $(\alpha L_P)^4$ where 
   $\alpha$ is a numerical constant. Then  ${\cal V}=N(\alpha L_P)^4$, implying a Poisson fluctuation $\Delta{\cal V}\approx
\sqrt{{\cal V}}(\alpha L_P)^2$  and leading to 
\begin{equation}
\Delta\Lambda={8 \pi L_P^2\over \Delta{\cal V}}= \left({8\pi \over \alpha^2}\right){1\over\sqrt{{\cal V}}}\approx  {8\pi \over \alpha^2} H_0^2
\label{dellamb}
\end{equation}
This will give $\Omega_\Lambda= (8\pi/3\alpha^2)$ which will --- for example --- lead to $\Omega_\Lambda =(2/3)$ if $\alpha = 2 \sqrt{\pi}$. Thus
Planck length cutoff (UV limit) and volume of the universe (IR limit) combine to give the correct $\Delta\Lambda$. 

 A similar result was obtained earlier by Sorkin \cite{sorkin} based on a different model. The numerical result can of course arise in different contexts and it is probably worth discussing some of the conceptual components in my argument.  The  key idea, in this approach,  is that $\Lambda$ is a stochastic variable with a zero mean and fluctuations. It is the rms fluctuation which is being observed in the cosmological context. This has 
 three implications: First, FRW equations now need to be solved with a stochastic term on the right hand side and one should check whether the observations can still be explained.
 The second  feature is that stochastic properties of $\Lambda$ need to be described by a quantum cosmological model. If the quantum state of the universe is expanded in terms of the eigenstates of some suitable operator (which does not commute the total four volume operator), then one should be able to characterize the fluctuations in each of these states.  Third, and most important, the idea of cosmological constant arising as 
 a {\it fluctuation} makes sense only if the bulk value is rescaled away; I have provided a toy model showing how this could be done. [In contrast, \cite{sorkin}, for example, {\it assumes }
the bulk value to be zero.]

 To show the nontriviality of this result, let me compare it with few other alternative 
 ways of estimating the fluctuations --- none of which gives the correct result. The first
 alternative approach is based on the assumption
 that one can associate an entropy $S=(A_H/4L_P^2)$ with compact space
time horizons of area $A_H$.  One popular interpretation of this result is that horizon areas are quantized
in units of $L_P^2$ so that $S$ is proportional to the number of bits of information contained in the horizon area.
In this approach, horizon areas can be expressed in the form $A_H=A_P N$ where $A_P\propto L_P^2$
is a quantum of area and $N$ is an integer. Then the {\it fluctuations} in the area will be 
$\Delta A_H=A_P\sqrt{N}=\sqrt{A_PA_H}$. Taking $A_H\propto \Lambda^{-1}$ 
for the De Sitter horizon, we find that $\Delta\Lambda\propto H^2(HL_P)$ which is a 
lot smaller than what one needs.
Further, taking $A_H\propto r_H^2$, we find that $\Delta r_H\propto L_P$; that is, this result  essentially arises from the idea that the radius of the horizon is uncertain within one Planck length. This is quite true, of course, but
does not lead to large enough fluctuations. 

A more sophisticated way of getting this (wrong) result is to
relate the fluctuations in the cosmological constant
 to that of the volume of the universe is by using a canonical ensemble description for universes
 of proper Euclidean 4-volume \cite{hawk}. 
 Writing $V\equiv {\cal V}/8\pi L_P^2$ and treating
 $V$ and $\Lambda$ as the relevant variables, one can write a partition function for the 
 4-volume as 
 \begin{equation}
 Z(V) = \int_0^\infty g(\Lambda) e^{-\Lambda V} d\Lambda
 \end{equation}
 Taking the analogy with standard statistical mechanics (with the correspondence $V \to \beta$ and
 $\Lambda \to E$), we can evaluate the fluctuations in the cosmological constant in exactly the 
 same way as energy fluctuations in canonical ensemble. 
 (This is done in several standard text books; see, for example, \cite{tpvol1} 
 p. 194.)
 This will give
 \begin{equation}
 \left( \Delta \Lambda\right)^2 ={C\over V^2}; \qquad C = {\partial \Lambda\over \partial (1/V)} = - V^2 {\partial \Lambda\over \partial V}
 \label{fluct}
 \end{equation}
 where $C$ is the analogue of the specific heat. 
 Taking the 4-volume of the universe to be ${\cal V} = b H^{-4}=9b\Lambda^{-2}$ where $b$ is a numerical
 factor and using 
 $V = ({\cal V}/8\pi  L_P^2) $
we get $\Lambda \propto L_P^{-1} V^{-1/2}$.  It follows from (\ref{fluct}) that 
 \begin{equation}
 (\Delta \Lambda)^2 = {C\over V^2}={12\pi\over b} (L_PH^3)^2
 \label{canens}
 \end{equation}
 In other words $\Delta \Lambda\propto H^2(HL_P)$, which is the same result from area quantization and is
  a lot smaller 
 than the cosmologically significant value.

Interestingly enough, one could do slightly better by assuming that
the horizon {\it radius} is quantized in units of  Planck length, so that $r_H=H^{-1}=NL_P$. This will lead to the
fluctuations $\Delta r_H=\sqrt{r_HL_P}$ and using $r_H=H^{-1}\propto \Lambda^{-1/2}$, we get 
$\Delta\Lambda\propto H^2(HL_P)^{1/2}$ --- larger than (\ref{canens}) but still inadequate.  These conclusions stress, among other things, the difference between {\it fluctuations} and the {\it mean values}. For, if one assumes that every patch of the universe with size $L_P$  contained an energy $E_P$, then a universe with characteristic size $H^{-1}$ will contain the energy
 $E=(E_P/L_P)H^{-1}$. The corresponding energy {\it density} will be $\rho_V=(E/H^{-3})=(H/L_P)^2$ which leads to the correct result. But, of course, we do not know why every length scale $L_P$ should contain an energy $E_P$ and --- more importantly --- contribute coherently  to give the total energy.
In summary,  the existence of two length scales $H^{-1}$ and $L_P$ allows different 
 results for $\Delta \Lambda$ depending on how exactly
 the fluctuations are characterized ($ \Delta V \propto \sqrt{N}, \Delta A \propto \sqrt{N} $ or
$ \Delta r_H \propto \sqrt{N}$). Hence the result obtained above in
 (\ref{dellamb}) is non trivial.
 
 As an aside, one could ask under what circumstance  the canonical 
 ensemble will lead to the correct fluctuations of the order $\Delta \Lambda =3\Omega_\Lambda H^2$
 where $\Omega_\Lambda \approx (0.65 - 0.7)$ is a numerical factor.
 Using the statistical formula $(\Delta \Lambda)^2 = CV^{-2} = -(\partial \Lambda/\partial V)$
 and the relation $(\Delta \Lambda)^2 =3\Omega_\Lambda H^2$ we get
 \begin{equation}
 (\Delta \Lambda)^2 = -{\partial \Lambda\over \partial V} \propto H^4 \propto {\cal V}^{-1} \propto L_P^{-2} V^{-1}=kL_P^{-2} V^{-1}
 \end{equation}
  where $k=(9b/8\pi)\Omega_\Lambda^2$ is a known numerical constant of order unity. Integrating, we find 
  $\Lambda = kL_P^{-2} \ln (V_0/V)$ leading to 
  \begin{equation}
  {\cal V} ={\cal V}_0 \exp \left( - {L_P^2\Lambda\over k}\right)
  \label{volchange}
  \end{equation}
  Thus, to produce the correct fluctuations, the 4-volume of the universe need to
  decrease exponentially with $\Lambda$ while we normally expect ${\cal V} \propto \Lambda^{-2}$
  which is just a power law decrease. 
   It is rather curious  that our toy scalar field model leads to exactly the same relationship
   as in (\ref{volchange}). Note that the 
  rescaling obtained in (\ref{eqsix}) which was needed to cancel the bulk cosmological constant,
  changes the 4-volume from ${\cal V}_0$ to ${\cal V}$ where 
  \begin{equation}
  {\cal V}_0 \to {\cal V} ={\cal V}_0 f^4 = {\cal V}_0 q^{-4} = {\cal V}_0 \exp \left( - {L_P^2\Lambda\over
  2\pi}\right)
  \end{equation}
  This is precisely the dependence of ${\cal V}$ on the bulk value of the $\Lambda$ which is required to
  produce the correct fluctuations even in the canonical ensemble picture.
  [The  connection between these two approaches 
  is  not clear.
  If one takes the numerical coefficient seriously, then setting $k~=~2\pi$ gives
  $\Omega_\Lambda =(4\pi/3\sqrt{b})$.] 
  
 While I am not optimistic about the {\it details} of the  model suggested here, I find it attractive to think of the observed cosmological constant as arising from quantum {\it fluctuations} of some energy density rather than from bulk energy density. This is relevant in the context of standard discussions of the contribution of 
 zero-point energies to cosmological constant. I would expect the correct theory to regularise the divergences and make the zero point energy finite and about $L_P^{-4}$. This contribution is most likely to modify the microscopic structure of spacetime (e.g if the spacetime is naively thought of as due to stacking of Planck scale volumes, this will modify the stacking or shapes of the volume elements) and will not affect the bulk gravitational field when measured at scales coarse grained over sizes much bigger than the Planck scales. 
 
 I thank K.Subramanian for useful discussions.


\section*{References}

\begin{thebibliography}{99}

\bibitem{sperl} S. Perlmutter etal., {\it Astrophys. J}, {\bf 517} (1999) 565.
\bibitem{agr} A.G. Riess etal., {\it Astron. J}, {\bf 116} (1998) 1009.

\bibitem{yz} Y.Zeldovich, Pis'ma(Letters), Zh.Eksp.Teoret. Fiz,{\bf 6},1050 (1967).
\bibitem{dolgov} A. Dolgov, Phys.Rev. {\bf D55} (1997) 5881 [astro-ph/9608175]; astro-ph/9708045.
\bibitem{carol} S.M. Carroll, ``The Cosmological Constant'', {\it Living Rev. Relativity}, {\bf 4}, (2001),
1. , http://www.livingreviews.org/Articles/Volume4/2001-1carroll/.
\bibitem{sorkin} D. Sorkin, 1997 {\it Forks in the road, on the way to quantum gravity}, gr-qc/9706002.
\bibitem{hawk} See for example,  Hawking S.W. (1979) in {\it General Relativity - 
An Einstein centenary survey},  Eds. S.W. Hawking
and W. Israel, Cambridge University Press, Cambridge, 1979.
\bibitem{tpvol1} T. Padmanabhan, {\it Theoretical Astrophysics, volume I: Astrophysical Processes},
(2000), Cambridge University Press, Cambridge, p. 194.

\end{thebibliography}
\end{document}